\documentstyle[aps,pra,amsfonts,floats,epsfig,eqsecnum]{revtex}

\newcommand{\raus}{{\text{out}}}
\newcommand{\rein}{{\text{in}}}

\newcommand{\eins}{{\openone}}

\newcommand{\dt}{d t}
\newcommand{\domega}{d \omega}
\newcommand{\domegas}{d \omega'}

\newcommand{\tr}{\text{tr}\,}

\newcommand{\freq}[1]{{#1}}

\newcommand{\rest}[1]{{#1}}
\newcommand{\deter}[1]{\Vert #1 \Vert}
\newcommand{\langlerad}{\langle}
\newcommand{\ranglerad}{\rangle}
\newcommand{\langlemeso}{\prec}
\newcommand{\ranglemeso}{\succ}

\newcommand{\neu}{_{\text{exc}}}
\newcommand{\oI}{\bar{I}}
\newcommand{\Iin}{I_0}
\newcommand{\mIn}{{m_0}}

\newcommand{\mr}{r}
\newcommand{\mt}{t}
\newcommand{\mrp}{r'}
\newcommand{\mtp}{t'}
\newcommand{\noise}{{\cal F}}
\newcommand{\zeit}{\tau}
\newcommand{\Pex}{P_{\text{exc}}}
\newcommand{\Pin}{P_0}

\newcommand{\thres}{\text{th}}

\newcommand{\typ}{\text{typ}}
\newcommand{\cotan}{\,\text{cotan}\,}
\newcommand{\cotanh}{\,\text{cotanh}\,}
\newcommand{\projekt}{{\cal P}}
\newcommand{\stern}{*}

\begin{document}

\draft

\date{January 1999}
\title{Excess noise for coherent radiation propagating through
amplifying random media}
\author{M. Patra and C. W. J. Beenakker}
\address{Instituut-Lorentz, Universiteit Leiden, P.O. Box 9506, 2300 RA
Leiden, The Netherlands}

\twocolumn[
\widetext
\begin{@twocolumnfalse}

\maketitle

\begin{abstract}
A general theory is presented for the photodetection statistics of
coherent radiation that has been amplified by a disordered medium. The
beating of the coherent radiation with the spontaneous emission
increases the noise above the shot-noise level. The excess noise is
expressed in terms of the transmission and reflection matrices of the
medium, and evaluated using the methods of random-matrix
theory. Inter-mode scattering between $N$ propagating modes increases
the noise figure by up to a factor of $N$, as one approaches the laser
threshold. Results are contrasted with those for an absorbing medium.
\end{abstract}

\pacs{PACS numbers: 42.50.Ar, 42.25.Bs, 42.25.Kb, 42.50.Lc}

\vspace{0.5cm}

\narrowtext

\end{@twocolumnfalse}
]

\section{Introduction}

The coherent radiation emitted by a laser has a noise spectral density
$P$ equal to the time-averaged photocurrent $\oI$. This noise is
called photon shot noise, by analogy with electronic shot noise in
vacuum tubes. If the radiation is passed through an amplifying medium,
$P$ increases more than $\oI$ because of the excess noise due to
spontaneous emission~\cite{henry:96a}.  For an ideal
linear amplifier, the (squared) signal-to-noise ratio 
$\oI^2/P$ drops by
a factor of two as one increases the gain.
One says that the amplifier
has a noise figure of $2$.
This is a lower bound on the excess noise for a linear
amplifier~\cite{caves:82a}.

Most calculations of the excess noise assume that the amplification
occurs in a single propagating mode. (Recent examples include work by
Loudon and his
group~\cite{jeffers:93a,matloob:97a}.) The
minimal noise figure of $2$ refers to this case. Generalisation to
amplification in a multi-mode waveguide is straightforward if there is
no scattering between the modes. The recent interest in amplifying
random media~\cite{wiersma:97a} calls for an extension of the theory
of excess noise to include inter-mode scattering. Here we present such
an extension.

Our central result is an expression for the probability distribution
of the photocount in terms of the transmission and reflection matrices
$\mt$ and $\mr$ of the multi-mode waveguide. (The noise power $P$ is
determined by the variance of this distribution.) Single-mode results
in the literature are recovered for scalar $\mt$ and $\mr$. In the
absence of any incident radiation our expression reduces to the known
photocount distribution for amplified spontaneous
emission~\cite{beenakker:98a}. We find that inter-mode scattering
strongly increases the excess noise, resulting in a noise figure that
is much larger than $2$.

We present explicit calculations for two types of geometries,
waveguide and cavity, distinguishing between photodetection in
transmission and in reflection.  We also discuss the parallel with
absorbing media. We use the method of random-matrix
theory~\cite{beenakker:97a} to obtain the required information on the
statistical properties of the transmission and reflection matrices of
an ensemble of random media. Simple analytical results follow
if the number of modes $N$ is large (i.e. for high-dimensional
matrices). Close to the laser threshold, the noise figure $\noise$
exhibits large sample-to-sample fluctuations, such that the ensemble
average diverges. We compute for arbitrary $N\ge2$ the distribution $p(\noise)$
of $\noise$ in the ensemble of disordered cavities, and show that 
$\noise=N$ is the most probable value. This is the
generalisation to
multi-mode random media of the single-mode result $\noise=2$ in the
literature.

\begin{figure}[b!]
\centering
\epsfig{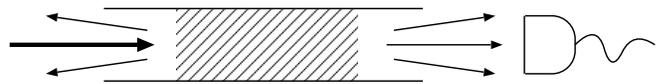}
\caption{Coherent light (thick arrow) is
incident on an amplifying medium (shaded),
embedded in a waveguide. The transmitted radiation is measured by a
photodetector.}
\label{aufbau}
\end{figure}

\section{Formulation of the problem}
\label{sectheorie}

We consider an amplifying disordered medium embedded in a waveguide
that supports $N(\omega)$ propagating modes at frequency $\omega$ (see
Fig.~\ref{aufbau}). The amplification could be due to stimulated
emission by an inverted atomic population or to stimulated Raman
scattering~\cite{henry:96a}. A negative temperature $T<0$ describes
the degree of population inversion in the first case or the density of
the material excitation in the second case~\cite{jeffers:93a}.  A
complete population inversion or vanishing density corresponds to the
limit $T\to0$ from below. The minimal noise figure mentioned in the
introduction is reached in this limit. The amplification rate
$1/\tau_a$ is obtained from the (negative) imaginary part $\epsilon''$
of the (relative) dielectric constant, $1/\tau_a=\omega |\epsilon''|$.
Disorder causes multiple scattering with rate $1/\tau_s$ and
(transport) mean free path $l=c\tau_s$ (with $c$ the velocity of light in
the medium). We assume that $\tau_s$ and $\tau_a$ are both
$\gg1/\omega$, so that scattering as well as amplification occur on
length scales large compared to the wavelength.
The waveguide is illuminated from one end by monochromatic radiation
(frequency $\omega_0$, mean photocurrent $\Iin$) in a coherent state. For
simplicity, we assume that the illumination is in a single
propagating mode (labelled $\mIn$). At the other end of the waveguide,
a photodetector detects the outcoming radiation. We assume, again for
simplicity, that all $N$ outgoing modes are detected with equal
efficiency $\alpha$.

We denote by $p(n)$ the probability to count $n$ photons within a time
$\zeit$. Its first two moments determine the mean photocurrent $\oI$ and
the noise power $P$, according to
\begin{equation}
\oI = \frac{1}{\zeit} \overline{n}, \qquad 
P = \lim_{\zeit\to\infty} \frac{1}{\zeit}\left(
\overline{n^2}-\overline{n}^2\right)\;.
	\label{groessen}
\end{equation}
(The definition of $P$ is equi\-val\-ent to
$P=\int_{-\infty}^{\infty} d t \,\overline{\delta I(0) \delta I(t)}$, 
\mbox{with $\delta I = I - \oI$} the
fluctuating part of the photocurrent.)
It is convenient to compute the generating function $F(\xi)$ for the factorial
cumulants $\kappa_j$, defined by
\begin{equation}
\label{Fxidef}
F(\xi)=\sum_{j=1}^{\infty} \frac{\kappa_j \xi^j}{j!}
	= \ln\left(\sum_{n=0}^{\infty} (1+\xi)^n p(n) \right)\;.
\end{equation}
One has $\overline{n}=\kappa_1, \overline{n^2} = \kappa_2 + \kappa_1
(1+\kappa_1)$.

The outgoing radiation in mode $n$ is described by an annihilation
operator $a_n^{\raus}(\omega)$, using the convention that modes
$1,2,\ldots,N$ are on the left-hand-side of the medium and modes
$N+1,\ldots,2 N$ are on the right-hand-side. The vector $a^{\raus}$
consists of the operators $a_1^{\raus},a_2^{\raus},\ldots,a_{2
N}^{\raus}$. Similarly, we define a vector $a^{\rein}$ for incoming
radiation. These two sets of operators each satisfy the bosonic
commutation relations
\begin{mathletters}
\label{commrel}
\begin{eqnarray}
[ a_n(\omega),a_m^\dagger(\omega')]&=&\delta_{nm}
	\delta(\omega-\omega')\;, \\
{[} a_n(\omega),a_m(\omega')] &=&0 \;,
\end{eqnarray}
\end{mathletters}
and are related by the
input-output relations~\cite{jeffers:93a,matloob:95a,gruner:96a}
\begin{equation}
        a^{\raus}(\omega) = S(\omega) a^{\rein}(\omega) +
        	V(\omega) c^\dagger(\omega) \;.
        \label{basiceq}
\end{equation}
We have introduced
the $2N\times2N$ scattering matrix $S$, the $2N\times2N$ matrix
$V$, and the vector $c$ of $2 N$ bosonic operators. The scattering matrix $S$
can be decomposed into four $N\times N$ reflection and 
transmission matrices,
\begin{equation}
	S = \left( \begin{array}{cc} \mr' & \mt' \\ \mt & \mr \end{array} \right) \;.
\end{equation}
Reciprocity imposes the conditions $\mt' = \mt^T$, $\mr = \mr^T$,
and $\mr'=\mr'^T$.

The operators $c$ account for spontaneous emission in the amplifying
medium. They satisfy the bosonic commutations
relation~(\ref{commrel}), which implies that
\begin{equation}
	V V^\dagger = S S^\dagger - \eins \;. 	\label{vvss}
\end{equation} 
Their expectation values are
\begin{equation}
\langlerad c_n(\omega) c_m^\dagger(\omega')\ranglerad
	= -\delta_{nm} \delta(\omega-\omega') f(\omega,T) \;,
	\label{cexpval}
\end{equation}
with the Bose-Einstein function
\begin{equation}
f(\omega,T)=\left[\exp(\hbar\omega/k T)-1\right]^{-1}
\end{equation}
evaluated at negative temperature $T$ ($<0$).


\section{Calculation of the generating function}
\label{sectgenerating}
\label{secF}

The probability $p(n)$ that $n$ photons are counted in a time $\zeit$ is given
by~\cite{glauber:63a,kelley:64a}
\begin{equation}
p(n) = \frac{1}{n!} \langlerad : W^n e^{-W} : \ranglerad \;,
\end{equation}
where the
colons denote normal ordering with respect to $a^{\raus}$, and
\begin{eqnarray}
&&  W = \alpha \int_0^\zeit 
	\dt \sum_{n=N+1}^{2N} a_n^{\raus\dagger}(t)
        	a_n^{\raus}(t)\;,
        \label{Wanfang} \\
&&  a^{\raus}_n(t)=(2\pi)^{-1/2} \int_0^\infty \text{d}\omega\,
        		e^{-i\omega t} a^\raus_n(\omega) \;. \label{avont}
\end{eqnarray}
The generating function
(\ref{Fxidef}) becomes
\begin{equation}
	F(\xi) = \ln\langlerad : e^{\xi W} : \ranglerad \;.	
	\label{Fxicomp}
\end{equation}

Expectation values of a normally ordered
expression are readily computed using the optical equivalence
theorem~\cite{mandel:95}. Application of this theorem to our problem
consists in discretising the frequency in infinitesimally small steps
of $\Delta$ (so that $\omega_p = p \Delta$) and then replacing the
annihilation operators $a^\rein_n(\omega_p), c_n(\omega_p)$ by complex
numbers $a^\rein_{np}$, $c_{np}$ (or their complex conjugates for the
corresponding creation operators). The coherent state of the incident
radiation corresponds to a non-fluctuating value of $a_{np}^\rein$
with $|a_{np}^\rein|^2=\delta_{n\mIn}\delta_{p p_0} 2 \pi \Iin /
\Delta$ (with $\omega_0=p_0\Delta$).
The thermal
state of the spontaneous emission corresponds to uncorrelated Gaussian
distributions of the real and imaginary parts of the numbers $c_{np}$,
with zero mean and variance $\langlerad (\text{Re}\,c_{np})^2
\ranglerad = \langlerad (\text{Im}\,c_{np})^2 \ranglerad=-\frac{1}{2}
f(\omega_p,T)$. (Note that $f<0$ for $T<0$.) To evaluate the
characteristic function (\ref{Fxicomp}) we need to perform 
Gaussian averages. The calculation is described in the appendix.

The result takes a simple form in the long-time regime $\omega_c
\zeit\gg1$, where $\omega_c$ is the frequency within which $S(\omega)$
does not vary appreciably. We find
\begin{eqnarray}
&&F(\xi) = F\neu(\xi)\nonumber\\
&&   	\quad\mbox{}-\frac{\zeit}{2\pi} \int_0^\infty \ln
	\deter{ \eins - \alpha \xi f ( \eins - \mr\mr^\dagger - \mt\mt^\dagger
		) } \,\domega\;, \label{altergebnis}\\
&&F\neu(\xi) =
       \alpha\xi \zeit \Iin \nonumber\\
&&       \quad{}\times\left( \mt^\dagger
       \left[ \eins - \alpha\xi f ( \eins - \mr \mr^\dagger - \mt \mt^\dagger )
       \right]^{-1} \mt\right)_{\mIn\mIn}
	\;,
	\label{LONGTIMEEINFACH}
	\label{hauptergebnis}
\end{eqnarray}
where $\deter{\cdots}$ denotes the determinant and $(\cdots)_{\mIn\mIn}$ the
$\mIn{,}\mIn$-element of a matrix. In Eq.~(\ref{hauptergebnis})
the functions $f$,
$\mt$, and $\mr$ are to be evaluated at $\omega=\omega_0$.
The integral in Eq.~(\ref{altergebnis}) is the
generating function for the photocount due to amplified spontaneous
emission obtained in Ref.~\onlinecite{beenakker:98a}. It is independent of
the incident radiation and can be eliminated in a measurement by
filtering the output through a narrow frequency window around
$\omega_0$. The function $F\neu(\xi)$ describes the excess noise due
to the beating of the coherent radiation with the spontaneous
emission. The expression (\ref{hauptergebnis}) is the central result
of this paper.

By expanding $F(\xi)$ in powers of $\xi$ we obtain the factorial
cumulants, in view of Eq.~(\ref{Fxidef}). We will in what follows
consider only the contribution from $F\neu(\xi)$, assuming that the
contribution from the integral over $\omega$ has been filtered out as mentioned
above. We find
\begin{equation}
\kappa_k = k! \alpha^k \zeit 
        f^{k-1}\Iin
        [\mt^\dagger
        \left(\eins-\mr \mr^\dagger - \mt \mt^\dagger \right)^{k-1}
	\mt]_{\mIn\mIn} \;,
\end{equation}
where again $\omega=\omega_0$ is implied. The mean photocurrent
$\oI=\kappa_1/\zeit$ and the noise power $P=(\kappa_2+\kappa_1)/\zeit$
become
\begin{eqnarray}
&& \oI = \alpha \Iin
        \left(\mt^\dagger \mt\right)_{\mIn\mIn}\;,\quad
	 P=\oI + \Pex\;,\nonumber\\
&& 	\Pex = 2 \alpha^2 
        f\Iin
        [\mt^\dagger
        \left(\eins-\mr \mr^\dagger - \mt \mt^\dagger \right)
	\mt]_{\mIn\mIn} \;.
\end{eqnarray}
The noise power $P$ exceeds the shot noise $\oI$ by the amount
$\Pex$. 

The formulas above are easily adapted to a measurement in
reflection by making the exchange $r\rightarrow t'$, $t\rightarrow r'$.
For example, the mean reflected photocurrent is
$ \oI = \alpha \Iin
        \left(\mrp^\dagger \mrp\right)_{\mIn\mIn}$, 
while the excess noise is
\begin{equation}	 
\Pex = 2 \alpha^2 
        f\Iin
        [\mrp^\dagger
        \left(\eins-\mrp \mrp^\dagger - \mtp \mtp^\dagger \right)
	\mrp]_{\mIn\mIn} \;.
\end{equation}


\section{Noise figure}

The noise figure $\noise$ is defined as the (squared) signal-to-noise ratio at
the input $\Iin^2/\Pin$, divided by the signal-to-noise ratio at the output,
$\oI^2/P$. Since $\Pin=\Iin$ for coherent radiation at the input, one has
$\noise = (P\neu+\oI)\Iin /\oI^2$, hence
\begin{equation}
\noise	= - 2 f \frac{ \left(\mt^\dagger \mr \mr^\dagger \mt
		+ \mt^\dagger \mt \mt^\dagger \mt\right)_{\mIn\mIn}}{
		\left( \mt^\dagger \mt\right)_{\mIn\mIn}^2}
	+ \frac{1+2\alpha f}{\alpha \left( \mt^\dagger \mt\right)_{\mIn\mIn}}
	\;.
	\label{noiseformel1}
\end{equation}	
The noise figure is independent of $\Iin$. For large amplification the second
term on the right-hand-side can be neglected relative to the first, and the
noise figure becomes also independent of the detection efficiency $\alpha$. The
minimal noise figure for given $r$ and $t$ is reached for an ideal detector
($\alpha=1$) and at complete population inversion ($f=-1$).

Since $(\mt^\dagger \mr \mr^\dagger \mt
		+ \mt^\dagger \mt \mt^\dagger \mt)_{\mIn\mIn}=\sum_k
		|(\mt^\dagger \mr)_{\mIn k}|^2+\sum_k |(\mt^\dagger
		\mt)_{\mIn k}|^2 \ge (\mt^\dagger \mt)_{\mIn\mIn}^2$,
one has $\noise\ge-2 f$ for large amplification (when the second term on the
right-hand-side of Eq.~(\ref{noiseformel1}) can be neglected). The minimal noise
figure $\noise=2$ at complete population inversion is reached in the absence of
reflection [$(\mt^\dagger \mr)_{\mIn k}=0$] and in the absence of inter-mode
scattering [$(\mt^\dagger \mt)_{\mIn k}=0$ if $k\ne\mIn$]. This is realised in
the single-mode theories of Refs.~\onlinecite{jeffers:93a,matloob:97a}. Our
result~(\ref{noiseformel1}) generalises these theories to include scattering
between the modes, as is relevant for a random medium.
		
These formulas apply to detection in transmission. For detection in reflection
one has instead
\begin{equation}
\noise 
	= - 2 f \frac{ \left(\mrp^\dagger \mtp \mtp^\dagger \mrp
		+ \mrp^\dagger \mrp \mrp^\dagger \mrp\right)_{\mIn\mIn}}{
		\left( \mrp^\dagger \mrp\right)_{\mIn\mIn}^2}
	+ \frac{1+2\alpha f}{\alpha \left( \mrp^\dagger \mrp\right)_{\mIn\mIn}}
	\;.
	\label{noiseformel2}
\end{equation}	
Again, for large amplification the second term on the right-hand-side may be
neglected relative to the first. The noise figure then becomes smallest in the
absence of transmission, when $\noise = - 2 f (\mrp^\dagger \mrp \mrp^\dagger
\mrp)_{\mIn\mIn}  (\mrp^\dagger \mrp)_{\mIn\mIn}^{-2}\ge- 2f$. The minimal noise
figure of $2$ at complete population inversion requires $(\mrp^\dagger \mrp
\mrp^\dagger \mrp)_{\mIn\mIn} = (\mrp^\dagger \mrp)_{\mIn\mIn}^2$, which is possible
only in the absence of inter-mode scattering.

To make analytical progress in the evaluation of $\noise$, we will
consider an ensemble of random media, with different realisations of
the disorder. For large $N$ and away from the laser threshold,
the sample-to-sample fluctuations in
numerators and denominators of Eqs.~(\ref{noiseformel1}) and
(\ref{noiseformel2}) are small, so we may average them
separately. 
Furthermore, the ``equivalent channel approximation'' is
accurate for random media~\cite{mello:92a}, which says that the
ensemble averages are independent of the mode index $\mIn$. Summing
over $\mIn$, we may therefore write $\noise$ as the ratio of traces,
so the noise figure for a measurement in transmission becomes
\begin{equation}
	\noise = -2 f N \frac{\langlemeso \tr ( \mt^\dagger \mr \mr^\dagger
	\mt + \mt^\dagger \mt \mt^\dagger \mt)\ranglemeso}{\langlemeso \tr 
	\mt^\dagger \mt \ranglemeso^2} +N\frac{1+2\alpha f}{\alpha \langlemeso
	\tr \mt^\dagger \mt\ranglemeso}\;,
	\label{traceequation}
\end{equation}
and similarly for a measurement in reflection. The brackets
$\langlemeso \cdots \ranglemeso$ denote the ensemble average.


\section{Applications}


\begin{figure}
\vspace{1.85mm}
\epsfig{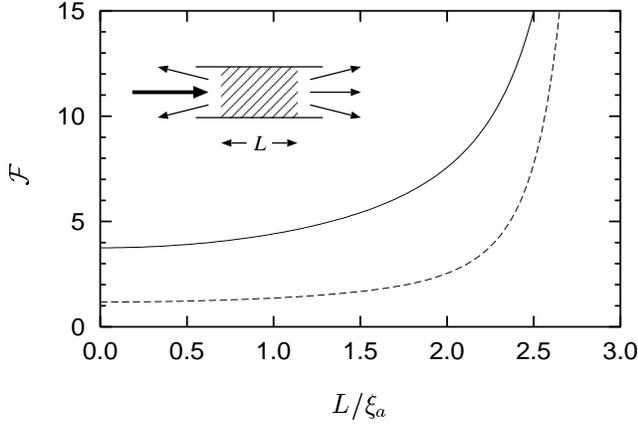}
\vspace{0.0mm}
\caption{Noise figure of an amplifying disordered waveguide (length
$L$, amplification length $\xi_a$) measured in transmission
(solid line) and 
in reflection (dashed line).
The curves are computed from Eqs.~(\ref{eqliste1a})--(\ref{eqliste1e})
for $\alpha=1$, $f=-1$, and $L/l=10$. The laser
threshold is at $L/\xi_a=\pi$.
}
\label{slab6fig}
\end{figure}

\subsection{Amplifying disordered waveguide}

As a first example, we consider a weakly amplifying,
strongly disordered waveguide of length $L$ (see inset of
Fig.~\ref{slab6fig}). Averages of the
moments of $\mr \mr^\dagger$ and $\mt \mt^\dagger$
for this system have been computed by
Brouwer~\cite{brouwer:98a} as a function of the number of
propagating modes $N$, the mean
free path $l$, and the amplification length
$\xi_a=\sqrt{D \tau_a}$, where $1/\tau_a$ is the
amplification rate and $D=c l/3$ is the diffusion constant. 
It
is assumed that $1/N \ll l/\xi_a \ll 1$ but the ratio $L/\xi_a\equiv s$ is
arbitrary. In this regime, sample-to-sample fluctuations are small,
so the ensemble average is representative of a single system.

The results for a measurement in transmission are 
\begin{eqnarray}
\oI & = & \frac{4 \alpha l}{3 L} \Iin
        \frac{s}{\sin s} \;, \label{tempeq4}\label{eqliste1a} \\
\Pex & = &
        \frac{2\alpha^2 l}{3 L} f \Iin s \left[
        \frac{3}{\sin s}-\frac{2 s - \cotan s}{\sin^2 s} \right.\nonumber\\
&&       \mbox{}\left.\quad\mbox{} + \frac{s \cotan s - 1}{\sin^3 s}
        - \frac{s}{\sin^4 s}
        \right] \;.
\end{eqnarray}
For a measurement in reflection, one finds
\begin{eqnarray}
\oI & = & \alpha \Iin \left[ 1 - \frac{4 l}{3 L}
        s \cotan s \right] \;,\\
\Pex & = &
        \frac{2\alpha^2 l}{3 L} f \Iin s \left[
        2 \cotan s
        - \frac{1}{\sin s}
        + \frac{\cotan s}{\sin^2 s} \right.\nonumber\\
&&	{}\left.\quad\mbox{}+ \frac{s \cotan s-1}{\sin^3 s}
        - \frac{s}{\sin^4 s}
        \right] \label{eqliste1e}\;.
\end{eqnarray}
The noise figure $\noise$ follows from $\noise =(\Pex+\oI) \Iin /
\oI^2$. It is plotted in Fig.~\ref{slab6fig}. One notices a strong
increase in $\noise$ on approaching the laser threshold at $s=\pi$.


\begin{figure}
\epsfig{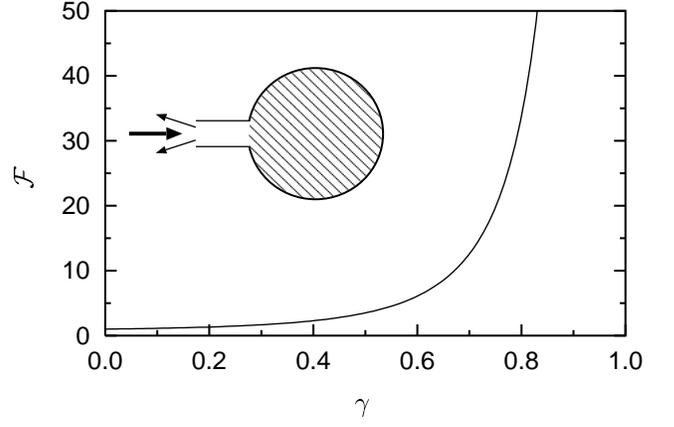}
\vspace{1.0mm}
\caption{Noise figure of an amplifying disordered cavity, connected to
a photodetector via an $N$-mode waveguide. The curve is the
result~(\ref{qdotnoiseformel}), as a function of the dimensionless
amplification rate $\gamma$. 
(Ideal detection efficiency, $\alpha=1$, and full population inversion,
$f=-1$, are assumed in this plot.)
The laser threshold occurs at $\gamma=1$.
}
\label{qdot2fig}
\end{figure}

\subsection{Amplifying disordered cavity}

Our second example is an optical cavity
filled with an amplifying random medium
(see inset of Fig.~\ref{qdot2fig}). The radiation leaves
the cavity through a waveguide supporting $N$ modes. The formulas
for a measurement in reflection apply with $\mt = 0$
because there is no transmission.
The distribution of the eigenvalues of $r^\dagger r$ is
known in the large-$N$ limit~\cite{beenakker:98b} as a function of the
dimensionless amplification rate $\gamma=2\pi/N\tau_a \Delta\omega$ (with
$\Delta\omega$ the spacing of the cavity modes near frequency
$\omega_0$). The first two moments of this
distribution are
\begin{eqnarray}
N^{-1} \langlemeso \tr r^\dagger r \ranglemeso
	&=&\frac{1}{1-\gamma}, \\
N^{-1} \langlemeso \tr r^\dagger r r^\dagger r \ranglemeso
	&=& \frac{2\gamma^2-2\gamma+1}{(1-\gamma)^4} \;.
\end{eqnarray}
The resulting photocurrent has mean and variance
\begin{eqnarray}
\oI & = & \alpha \Iin \frac{1}{1-\gamma}\;, 	\label{eqliste2a} \\
\Pex & = & 2 \alpha^2 f \Iin \gamma
	\frac{\gamma-\gamma^2-1}{(1-\gamma)^4}\;.
	\label{eqliste2e}
\end{eqnarray}
The resulting noise figure for $\alpha=1$ and $f=-1$,
\begin{equation}
	\noise =
	\frac{1-\gamma+\gamma^2+\gamma^3}{\left(1-\gamma\right)^2} \;,
	\label{qdotnoiseformel}
\end{equation}
is plotted in Fig.~\ref{qdot2fig}.
Again, we see a strong increase of $\noise$ on approaching the laser
threshold at $\gamma=1$.


\section{Near the laser threshold}

In the previous section we have taken the large-$N$ limit. In that
limit the noise figure diverges on approaching the laser threshold. In
this section we consider the vicinity of the laser threshold for
arbitrary $N$.

The scattering matrix $S(\omega)$ has poles in the lower half of the
complex plane. With increasing amplification, the poles shift
upwards. The laser threshold is reached when a pole reaches the real
axis, say at resonance frequency $\omega_{\thres}$. For $\omega$ near $\omega_{\thres}$
the scattering matrix has the generic form
\begin{equation}
	S_{nm} = \frac{\sigma_n
	\sigma_m}{\omega-\omega_{\thres}+\frac{1}{2} i \Gamma -
	i/2\tau_a} \;,
	\label{tempeq11}
\end{equation}
where $\sigma_n$ 
is the complex coupling constant of the resonance to
the $n$-th mode in the waveguide, $\Gamma$ is the decay rate, and
$1/\tau_a$ the amplification rate. The laser threshold is at
$\Gamma\tau_a=1$.

We assume that the incident radiation has frequency $\omega_0=\omega_{\thres}$.
Substitution of Eq.~(\ref{tempeq11}) into Eq.~(\ref{noiseformel1}) or
(\ref{noiseformel2}) gives the simple result
\begin{equation}
	\noise = \frac{- 2 f \Sigma}{|\sigma_{\mIn}|^2}, \;
	\Sigma = \sum_{n=1}^{2 N} |\sigma_n|^2 \;,
	\label{noiseallgemein}
\end{equation}
for the limiting value of the noise figure on approaching the laser threshold.
The limit is the same for detection in transmission and in reflection. Since
the coupling contant $|\sigma_{\mIn}|^2$ to the mode $\mIn$ of the incident
radiation can be much smaller than the total coupling constant $\Sigma$,
the noise figure (\ref{noiseallgemein}) has large fluctuations. We need to
consider the statistical distribution $p(\noise)$ in the ensemble of random
media. The typical (or modal) value of $\noise$ is the value $\noise_{\typ}$ at which
$p(\noise)$ is maximal. We will see that this remains finite although the
ensemble average $\langlemeso \noise \ranglemeso$ of $\noise$ diverges.


\subsection{Waveguide geometry}

We first consider the case of an amplifying disordered waveguide. The total
coupling constant $\Sigma=\Sigma_l+\Sigma_r$ is the sum of the coupling
constant $\Sigma_l=\sum_{n=1}^{N} |\sigma_n|^2$ to the left end of the
waveguide and the coupling constant $\Sigma_r=\sum_{n=N+1}^{2 N} |\sigma_n|^2$
to the right. The assumption of equivalent channels implies that
\begin{equation}
	\langlemeso 1/\noise \ranglemeso = - \frac{1}{2 f N}
	\langlemeso \Sigma_l / \Sigma \ranglemeso = - \frac{1}{4 f N}
	\;.
\end{equation}

Since the average of $1/\noise$ is finite, it is reasonable to assume that
$\noise_{\typ}\approx \langlemeso 1/\noise\ranglemeso^{-1}=-4 f N$, or
$\noise_{\typ}\approx 4 N$ for complete population inversion. The scaling with
$N$ explains why the large-$N$ theory of the previous section found a divergent
noise figure at the laser threshold. We conclude that the divergency
of $\noise$ at $L/\xi_a=\pi$ in Fig.~\ref{slab6fig}
is cut off at a value of order $N$, if $\noise$ is identified with
the typical value $\noise_{\typ}$.


\subsection{Cavity geometry}

In the case of an amplifying disordered cavity, we can make a more
precise statement on $p(\noise)$. Since there is only reflection there is only
one $\Sigma=\sum_{n=1}^{N} |\sigma_n|^2$. 
The assumption of equivalent channels now gives
\begin{equation}
	\langlemeso 1/\noise \ranglemeso = -\frac{1}{2 f N} \;.
	\label{tempeq10}
\end{equation}
Following the same reasoning as in the case of the waveguide, we would conclude
that $\noise_{\typ}\approx \langlemeso 1/\noise \ranglemeso^{-1}=-2 f N$. We
will see that this is correct within a factor of two.

To compute $p(\noise)$ we need the distribution of the dimensionless
coupling constants $u_n=\sigma_n/\sqrt{\Sigma}$. The $N$
complex numbers $u_n$ form a vector $\vec{u}$ of length $1$. According
to random-matrix theory~\cite{beenakker:97a}, the distribution $p(S)$ of
the scattering matrix is invariant under unitary transformations $S\to
U S U^T$ (with $U$ an $N\times N$ unitary matrix). It follows
that the distribution $p(\vec{u})$ of the vector $\vec{u}$ is
invariant under rotations $\vec{u}\to U \vec{u}$, hence
\begin{equation}
	p(u_1, u_2, \ldots, u_N) \propto \delta\left(1
		-\sum_n |u_n|^2 \right) \;.
\end{equation}
In other words, the vector $\vec{u}$ has the same distribution as a column of a
matrix that is uniformly distributed in the unitary group~\cite{pereyra:83a}.
By integrating out $N-1$ of the $u_n$'s we find the marginal
distribution of $u_\mIn$,
\begin{equation}
	p(u_\mIn) = \frac{N-1}{\pi} \left( 1 - |u_\mIn|^2 \right)^{N-2} \;,
\end{equation}
for $N\ge2$ and $|u_\mIn|^2\le 1$.

\begin{figure}
\epsfig{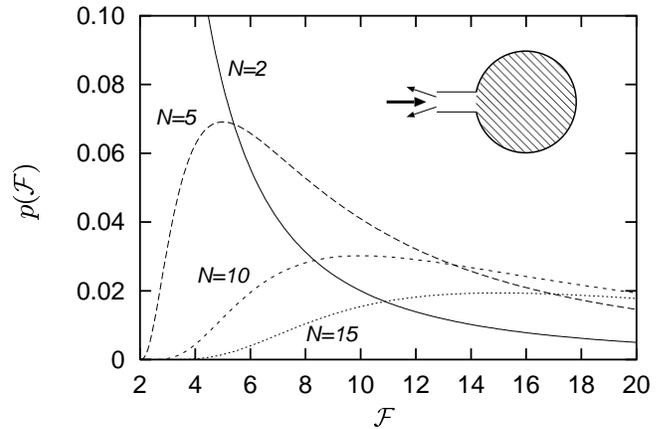}
\vspace{0.0mm}
\caption{Probability distribution of the noise figure 
near the laser threshold
for an amplifying
disordered cavity, computed from Eq.~(\ref{tempeq9}) for $f=-1$.
The most probable value is $\noise=N$, while the average value
diverges.
}
\label{noisedistribution}
\end{figure}

The distribution of $\noise = - 2 f |u_\mIn|^{-2}$ becomes
\begin{equation}
	p(\noise) =  - 2 f ( N - 1) \left( 1 + \frac{2 f}{\noise}
			\right)^{N-2} \noise^{-2} \;,
	\label{tempeq9}
\end{equation}
for $N\ge2$ and $\noise\ge-2 f$. We have plotted $p(\noise)$ in
Fig.~\ref{noisedistribution} for complete population inversion ($f=-1$) and
several choices of $N$. It is a broad distribution, all its moments are
divergent. The
typical value of the
noise figure is the value at which $p(\noise)$ becomes maximal, hence
\begin{equation}
	\noise_{\typ} = - f N, \quad N\ge 2 \;.
\end{equation}
In the single-mode case, in contrast, $\noise=-2 f$ for
every member of the ensemble  [hence $p(\noise)=\delta(\noise+2 f)$].
We conclude that the typical value of the noise figure near the laser threshold
of a disordered cavity is larger than in the single-mode case by a factor $N/2$.


\section{Absorbing media}

\begin{figure}[b!]
\epsfig{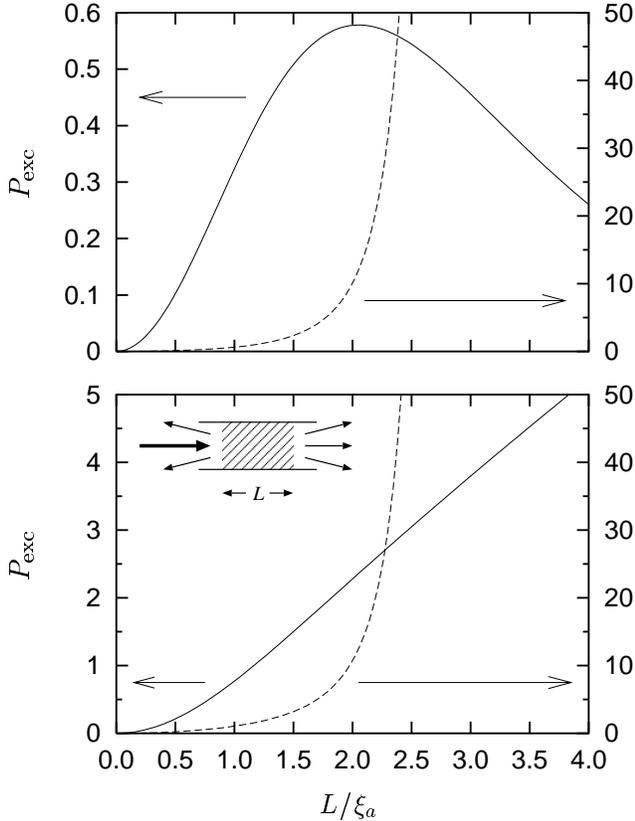}
\vspace{0.0mm}
\caption{Excess noise power $\Pex$ for an absorbing (solid
line, left axis) respectively amplifying disordered waveguide
(dashed line, right axis),
in units of $\alpha^2 l |f| \Iin / L$. The top panel is for
detection in transmission, the bottom panel for detection in
reflection.}
\label{slab12fig}
\end{figure}

The general theory of Sec.~\ref{sectheorie} can also be applied to 
an absorbing medium, in equilibrium at temperature $T>0$.
Eq.~(\ref{basiceq}) then has to be replaced with
\begin{equation}
        a^{\raus}(\omega) = S(\omega) a^{\rein}(\omega) +
        	Q(\omega) b(\omega) \;,
\end{equation}
where the bosonic operator $b$ has the expectation value
\begin{equation}
\langlerad b_n^\dagger(\omega) b_m(\omega')\ranglerad
	= \delta_{nm} \delta(\omega-\omega') f(\omega,T) \;,
\end{equation}
and the matrix $Q$ is related to $S$ by
\begin{equation}
        Q Q^\dagger = \eins - S S^\dagger \;.
\end{equation}        
The formulas for $F(\xi)$ of Sec.~\ref{secF} remain unchanged.

Ensemble averages for absorbing systems follow from the corresponding results
for amplifying systems by substitution $\tau_a\to-\tau_a$. 
(This follows from a
general duality theorem~\cite{paasschens:96a} between absorbing and amplifying
systems.) The results for an absorbing disordered waveguide with detection in
transmission are
\begin{eqnarray}
\oI & = &
        \frac{4 \alpha l}{3 L} \Iin
        \frac{s}{\sinh s}\;, \\
\Pex & = &
        \frac{2\alpha^2 l}{3 L} f \Iin s\left[
        \frac{3}{\sinh s}-\frac{2 s + \cotanh s}{\sinh^2 s} \right.\nonumber\\
&&      {}\left.\quad\mbox{}- \frac{s \cotanh s - 1}{\sinh^3 s}
	+ \frac{s}{\sinh^4 s}
        \right] \;,
\end{eqnarray}
where $s=L/\xi_a$ with $\xi_a$ the absorption length.
Similarly, for detection in reflection one has
\begin{eqnarray}
\oI & = &
        \alpha \Iin \left[ 1 - \frac{4 l}{3 L}
        s \cotanh s \right] \;,\\
\Pex & = &
        \frac{2\alpha^2 l}{3L} f \Iin s \left[
        2 \cotanh s
        - \frac{1}{\sinh s}- \frac{\cotanh s}{\sinh^2 s} \right.\nonumber\\
&&      {}\quad\left.
	\mbox{}- \frac{s \cotanh s-1}{\sinh^3 s}
        + \frac{s}{\sinh^4 s}
        \right] \;.
\end{eqnarray}
These formulas follow from Eqs.~(\ref{eqliste1a})--(\ref{eqliste1e})
upon substitution of $s\to i s$.

For an absorbing disordered cavity, we find [substituting
$\gamma\to-\gamma$ in Eqs.~(\ref{eqliste2a})--(\ref{eqliste2e})],
\begin{eqnarray}
\oI & = & \alpha \Iin \frac{1}{1+\gamma} \;,\\
\Pex & = & 2 \alpha^2 f \Iin \gamma
	\frac{\gamma^2+\gamma+1}{(1+\gamma)^4} \;,
\end{eqnarray}
with $\gamma$ the dimensionless absorption rate.

\begin{figure}
\epsfig{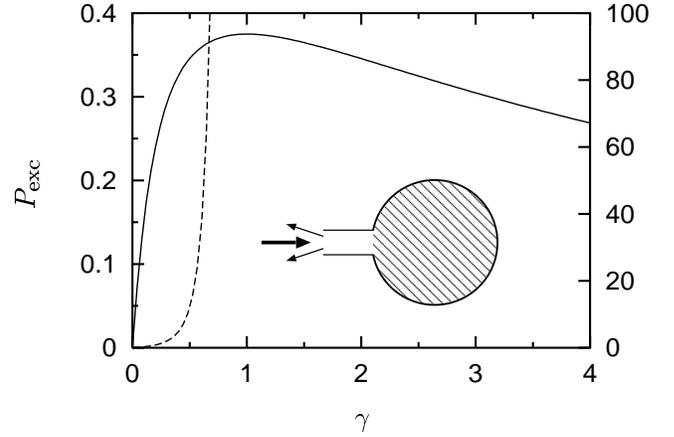}
\vspace{0.0mm}
\caption{Excess noise power $\Pex$ for an absorbing (solid line, left axis)
respectively amplifying disordered cavity (dashed line, right axis),
in units of $\alpha^2 |f| \Iin$.}
\label{qdot1fig}
\end{figure}

Since typically $f\ll 1$ in absorbing systems, the noise figure $\noise$ is
dominated by shot noise, $\noise\approx \Iin/\oI$. Instead of $\noise$ we
therefore plot the excess noise power $\Pex$ in Figs.~\ref{slab12fig}
and \ref{qdot1fig}.
In contrast to the
monotonic increase of $\Pex$ with $1/\tau_a$ in amplifying systems, the
absorbing systems show a maximum in $\Pex$ for certain geometries. The maximum
occurs near $L/\xi_a=2$ for the disordered waveguide with detection in
transmission, and near $\gamma=1$ for the disordered cavity. For larger
absorption rates the excess noise power decreases because $\oI$ becomes too
small for appreciable beating with the spontaneous emission.


\section{Conclusion}

In summary, we have studied the photodetection statistics of coherent
radiation that has been transmitted or reflected by an amplifying or
absorbing random medium. The cumulant generating function $F(\xi)$ is
the sum of two terms. The first term is the contribution from
spontaneous emission obtained in Ref.~\onlinecite{beenakker:98a}. The second
term $F_{\text{exc}}$ is the excess noise due to beating of the
coherent radiation with the spontaneous
emission. Equation~(\ref{LONGTIMEEINFACH}) relates $F_{\text{exc}}$
to the transmission and reflection matrices of the medium.

In the applications of our general result for the cumulant generating
function we have concentrated on the second cumulant, which gives the
spectral density $P_{\text{exc}}$ of the excess noise. We have found
that $P_{\text{exc}}$ increases monotonically with increasing
amplification rate, while it has a maximum as a function of absorption
rate in certain geometries.

In amplifying systems we studied how the noise figure $\noise$
increases on approaching the laser threshold. 
Near the laser threshold the noise figure shows large
sample-to-sample fluctuations, such that its statistical distribution
in an ensemble of random media has divergent first and
higher moments. The most probable value of $\noise$ is of the order of
the number $N$ of propagating modes in the medium, independent of material
parameters such as the mean free path. It would be of interest to observe
this universal limit in random lasers.

\acknowledgements

We thank P.\ W.\ Brouwer for helpful comments. This work was supported by
the Nederlandse Organisatie voor Wetenschappelijk Onderzoek (NWO) and
the Stichting voor Fundamenteel Onderzoek der Materie (FOM).


\appendix

\section*{Derivation of Eq.~(\ref{LONGTIMEEINFACH})}

To evaluate the Gaussian averages that lead to 
Eq.~(\ref{LONGTIMEEINFACH}), it is convenient to use a matrix notation.
We replace the summation in Eq.~(\ref{Wanfang}) by a multiplication of
the vector $a^\raus$ with the projection $\projekt a^\raus$, where the
projection matrix $\projekt$ has zero elements except
$\projekt_{nn}=1$, $N+1\le n\le 2 N$. We thus write
\begin{equation}
        W = \alpha \int_0^\zeit \dt \, a^{\raus\dagger}(t) \projekt
        	a^{\raus}(t)  \;.
\end{equation}
Insertion of Eqs.~(\ref{basiceq}) and (\ref{avont}) gives
\begin{eqnarray}
W &=& \frac{\alpha}{2\pi} \int_0^\zeit\!\!\dt
	\int_0^\infty\!\!\!\!\domega \int_0^\infty\!\!\!\!\domegas
	\!\left[ a^{\rein\dagger}(\omega) S^\dagger(\omega)  +
        	c(\omega) V^\dagger(\omega) \right] \nonumber \\
& & \times	\projekt 
	\left[ S(\omega') a^{\rein}(\omega') +
        	V(\omega') c^\dagger(\omega') \right]
	e^{i(\omega-\omega')t} \;. \label{temp34}
\end{eqnarray}
As explained in Sec.~\ref{sectgenerating} we discretise the frequency as
$\omega_p=p\Delta$, $p=1,2,3,\ldots$. 
The integral over frequency is then replaced with a summation,
\begin{equation}
	\int_0^\infty \domega \,g(\omega) \to
	\Delta \sum_{p=1}^\infty g(\omega_p)\;.
\end{equation}
We write Eq.~(\ref{temp34}) as a matrix multiplication,
\begin{equation}
\xi W = \freq{a}^{\rein\dagger} \rest{A} \freq{a}^{\rein}
                + \freq{c} \rest{B} \freq{c}^\dagger
                + \freq{a}^{\rein\dagger} \rest{C}^{\dagger} \freq{c}^\dagger
                + \freq{c} \rest{C} \freq{a}^{\rein} \;,
\end{equation}
with the definitions
\begin{eqnarray}                
        \rest{A}_{np,n'p'} & = &
	\frac{\alpha\Delta\xi}{2\pi} \int_0^\zeit \dt
		\left( S^\dagger(\omega_p) \projekt S(\omega_{p'})
		\right)_{nn'} e^{i\Delta (p-p')t}
		\;, \nonumber\\
        \rest{B}_{np,n'p'} & = & \frac{\alpha\Delta\xi}{2\pi}
		\int_0^\zeit \dt\left( V^\dagger(\omega_p) \projekt V(\omega_{p'})
                \right)_{nn'} e^{i\Delta(p-p')t} \;,\nonumber\\
        \rest{C}_{np,n'p'} & = & \frac{\alpha\Delta\xi}{2\pi} \int_0^\zeit\dt
        	\left( V^\dagger(\omega_p) \projekt S(\omega_{p'})
        	\right)_{nn'} e^{i\Delta(p-p')t}
                 \;,\nonumber\\
	\freq{a}^{\rein}_{np} & = & \Delta^{1/2} a^{\rein}_n(\omega_p), \qquad
	\freq{c}_{np} = \Delta^{1/2} c_n(\omega_p) \;.
		\label{matrixdefs}
\end{eqnarray}

We now apply the optical equivalence theorem~\cite{mandel:95}, as
discussed in Sec.~\ref{secF}. The operators $a^\rein_{np}$ are
replaced by constant numbers $\delta_{n\mIn} \delta_{p p_0} ( 2 \pi
\Iin / \Delta )^{1/2}$. The operators $c_{np}$ are replaced by
independent Gaussian variables, such that the expectation value
(\ref{Fxicomp})
takes the form of a Gaussian integral,
\begin{eqnarray}
\lefteqn{\langlerad : e^{\xi W} : \ranglerad =
        \int d\left\lbrace \freq{c}_{np}\right\rbrace
	\exp\left[ \xi W + \sum_{np} |c_{np}|^2 / f(\omega_p,T)\right]}
        &&\mbox{}\hspace{3.4in}\mbox{}\nonumber\\
\lefteqn{ = \int d\left\lbrace \freq{c}_{np}\right\rbrace
        \exp\left[ \freq{a}^{\rein\stern} \rest{A} \freq{a}^{\rein}
                - \freq{c} \rest{M} \freq{c}^{\stern}
                + \freq{a}^{\rein\stern} \rest{C}^{\dagger} \freq{c}^{\stern}
                + \freq{c} \rest{C} \freq{a}^{\rein}
        \right]\;,}&&\nonumber\\
	\label{Fxicomp2}
\end{eqnarray}
where we have defined
\begin{equation}
        \rest{M}_{np,n'p'} = -\rest{B}_{np,n'p'} - 
        \frac{\delta_{nn'}\delta_{pp'}}{f(\omega_p)} \;.
\end{equation}

We eliminate the cross-terms of $\freq{a}^{\rein}$ and
$\freq{c}$ in Eq.~(\ref{Fxicomp2}) by the substitution
\begin{equation}
	\freq{c'}^{\stern} = \freq{c}^{\stern} - \rest{M}^{-1}
	\rest{C} \freq{a}^{\rein} \;,
\end{equation}
leading to
\begin{eqnarray}
\langlerad : e^{\xi W} : \ranglerad & = &
        \exp\left[
        \freq{a}^{\rein\stern} ( \rest{A} +
        \rest{C}^{\dagger}
        \rest{M}^{-1} \rest{C} ) \freq{a}^{\rein}
        \right] \nonumber\\
& & \times
        \int d\left\lbrace \freq{c'}_{np}\right\rbrace
        \exp\left( -\freq{c'} \rest{M}
        \freq{c'}^{\stern} \right) \;.
\end{eqnarray}
The integral is proportional to
the determinant of $\rest{M}^{-1}$, giving the
generating function
\begin{eqnarray}
F(\xi) &=& \mbox{constant} - \ln \deter{\rest{M}}
        + \freq{a}^{\rein\stern} \left( \rest{A} 
        + \rest{C}^{\dagger} \rest{M}^{-1}
        \rest{C} \right)\freq{a}^{\rein} \nonumber\\
& = & \mbox{constant} - \ln \deter{\rest{M}}\nonumber\\
        && \mbox{} + \frac{2\pi\Iin}{\Delta}
	\left( \rest{A} + \rest{C}^{\dagger} \rest{M}^{-1}
        \rest{C} \right)_{\mIn p_0,\mIn p_0} \;.
        \label{fxierg}
\end{eqnarray}
The additive constant follows from $F(0)=0$.
The term $-\ln \deter{\rest{M}}$ is the contribution from amplified
spontaneous emission calculated in
Ref.~\onlinecite{beenakker:98a}. The term proportional to $\Iin$ is
the excess noise of the coherent radiation,
termed $F\neu$ in Sec.~\ref{secF}.


Eq.~(\ref{fxierg})
can be simplified in the long-time regime,
$\omega_c \zeit\gg 1$.
We may then set $\Delta=2\pi/\zeit$ and use
\begin{equation}
        \int_0^\zeit e^{i \Delta(p - p')t} \dt =
         \zeit \delta_{pp'} \;.
\end{equation}
The matrices defined in Eq.~(\ref{matrixdefs}) thus become diagonal in
the frequency index,
\begin{equation}
        \rest{A}_{np,n'p'} =
	\frac{\alpha\Delta\zeit\xi}{2\pi} 
		\left( S^\dagger(\omega_{p}) \projekt S(\omega_{p})
		\right)_{nn'} \delta_{pp'}
\end{equation}
and similarly for $\rest{B}$ 
and $\rest{C}$. We then find
\begin{eqnarray}
\lefteqn{(A + C^{\dagger} M^{-1} C)_{np,n'p'} =} \nonumber\\
& &     \mbox{~~~~~~~}\frac{\alpha\xi\Delta \zeit}{2\pi} (S^\dagger \projekt
        [ \eins + \alpha\xi f V V^\dagger \projekt ]^{-1} S)_{nn'} \delta_{pp'}\;,
\end{eqnarray}
where $f$, $S$, and $V$ are evaluated at $\omega=\omega_p$.
Substitution into Eq.~(\ref{fxierg}) gives the result
(\ref{LONGTIMEEINFACH}) for $F_{\text{exc}}(\xi)$.


Simplification of Eq.~(\ref{fxierg}) is also possible 
in the short-time regime, when
$\Omega_c \zeit\ll1$, with $\Omega_c$
the frequency range over which $S S^\dagger$ differs appreciably from
the unit matrix. The generating function then is
\begin{eqnarray}
&&F\neu(\xi)=\alpha\xi \zeit \Iin \biggl( \mt^\dagger(\omega_0) \Bigl[ \eins
	- \frac{\alpha\xi\zeit}{2\pi} \int_0^\infty \domega 
	f(\omega,T)\nonumber\\
\lefteqn{\times\left(\eins-\mr(\omega) \mr^\dagger(\omega)
	-\mt(\omega)\mt^\dagger(\omega)\right)\Bigr]^{-1}
	\mt(\omega_0)\biggr)_{\mIn\mIn}\!\!\!\;.}&&
\end{eqnarray}




\end{document}